\documentclass[aps,prl,preprint]{revtex4-2}

\usepackage[colorlinks=true, citecolor=blue, linkcolor=blue, urlcolor=blue]{hyperref}
\usepackage{graphicx}
\usepackage{amsmath}
\usepackage{soul} 
\usepackage{upgreek}

\begin{document}


\title{Two-Phonon Resonance Drives Multicomponent Mechanical Cat States}



\author{Nuo Wang$^{1}$}
\author{Haoyang Zhang$^{1,5}$}
\author{Yu Tian$^{1,2}$}
\author{Yadi Niu$^{1}$}
\author{Ying Gu$^{1,2,3,4,5,}$}
\email[]{ygu@pku.edu.cn}
\affiliation{$^1$State Key Laboratory of Artificial Microstructure and Mesoscopic Physics $\&$ Department of Physics, Peking University, Beijing 100871, China\\ $^2$Frontiers Science Center for Nano-optoelectronics $\&$  Collaborative Innovation Center of Quantum Matter $\&$ Beijing Academy of Quantum Information Sciences, Peking University, Beijing 100871, China\\$^3$Collaborative Innovation Center of Extreme Optics, Shanxi University, Taiyuan, Shanxi 030006, China\\$^4$Peking University Yangtze Delta Institute of Optoelectronics, Nantong 226010, China\\$^5$Hefei National Laboratory, Hefei 230088, China}


\date{\today}

\begin{abstract}
 Using quadratic optomechanical coupling to prepare high-purity mechanical cat states is not feasible as its strength is several orders weaker than linear optomechanical coupling. Here, using only linear  coupling in a multimode system, we achieve strong interaction between photons and two phonons, enabling the deterministic generation of high-purity multicomponent mechanical cats. Mediated by an auxiliary supermode, when other two optical supermodes satisfy the two-phonon resonance condition, the process  whereby the annihilation of a high-frequency photon  accompanied by the creation of a low-frequency photon and two phonons is strongly enhanced. Such resonant two-phonon process drives multiple rotations and interferences of mechanical coherent states, deterministically generating a multicomponent mechanical cat immune to both mechanical and optical losses.  Our work provides an universal strategy for enhancing high-order phonon nonlinearities,  paving the way for quantum state engineering, quantum precision measurement and fault-tolerant quantum computation.
 
\end{abstract}


\maketitle

\section{Introduction}

Nonlinear interactions in cavity optomechanical systems are essential for various quantum information processes, including non-Gaussian state preparation \cite{2016PRL.JQL,2023PRL.HAU}, quantum logic operations \cite{2012PRL.STA}, and quantum precision measurement \cite{2023PRL.CLA,2024OQ}.  Especially, two-phonon interaction between light and a mechanical oscillator plays a key role in generating two-component cat states \cite{2023PRL.HAU,2013PRA.TAN,2018PRA.BRU}, cubic phase states \cite{2019PRA}, and Wigner-negative entangled states \cite{2024PRA}. Typically, two-phonon interaction is realized via quadratic optomechanical coupling, which is proportional to the square of mechanical displacement \cite{2013PRA.TAN,2010PRA,2018PRA.BRU,2019PRA,2024PRA}. However, its strength is several orders weaker than that of linear optomechanical coupling \cite{2008NATURE,2012APL.FLO,2014PRA}. Alternatively, using linear optomecahnical coupling under bichromatic driving, an effective two-phonon interaction is proposed \cite{2023PRL.HAU}, but its strength is strongly suppressed in high-frequency systems.
 Consequently,  with above two-phonon interactions \cite{2013PRA.TAN,2010PRA,2018PRA.BRU,2019PRA,2024PRA,2023PRL.HAU}, it is difficult to  prepare high-purity non-Gaussian states  since  the interaction strength is  too weak to overcome the intrinsic mechanical and optical losses. 
 Thus,   one of our aims is to use only linear optomechanical coupling to achieve an enhanced two-phonon interaction with the strength approaching that of linear coupling. With such strong interaction, to generate important non-Gaussian mechanical quantum states is promising.

Mechanical cat states \cite{1935NATURW,1997AJP}, as typical non-Gaussian states, are of great significance for quantum precision measurement \cite{2002PRA,2017PRA}, fault-tolerant quantum computing \cite{2022PRXQ}, and fundamental tests of quantum mechanics \cite{1991PT,2013PRL.BLE}. There have been various schemes to generate two-component mechanical cat states, such as single-phonon subtraction \cite{2020PRA.SHO,2020PRA.ZHA}, coupling with a spin qubit \cite{2013PRA.YIN,2018PRL.SAN} and utilizing optomechanical nonlinearities \cite{2023PRL.HAU,2013PRA.TAN,2018PRA.BRU}. 
Beyond two-component  cat states, multicomponent cat states  have attracted  more interests due to their  greater capability in quantum error correction \cite{2014NJP,2017PRL.LI,2016NATURE,2021NATURE} and higher sensitivity in quantum metrology \cite{2001NATURE,2006NJP}. 
But in optomechanical systems, owing to the relatively weak nonlinear  interactions,  to generate multicomponent mechanical cat state (MMCS) still remains challenging.

\begin{figure}[t]
	\includegraphics[width=0.8\textwidth]{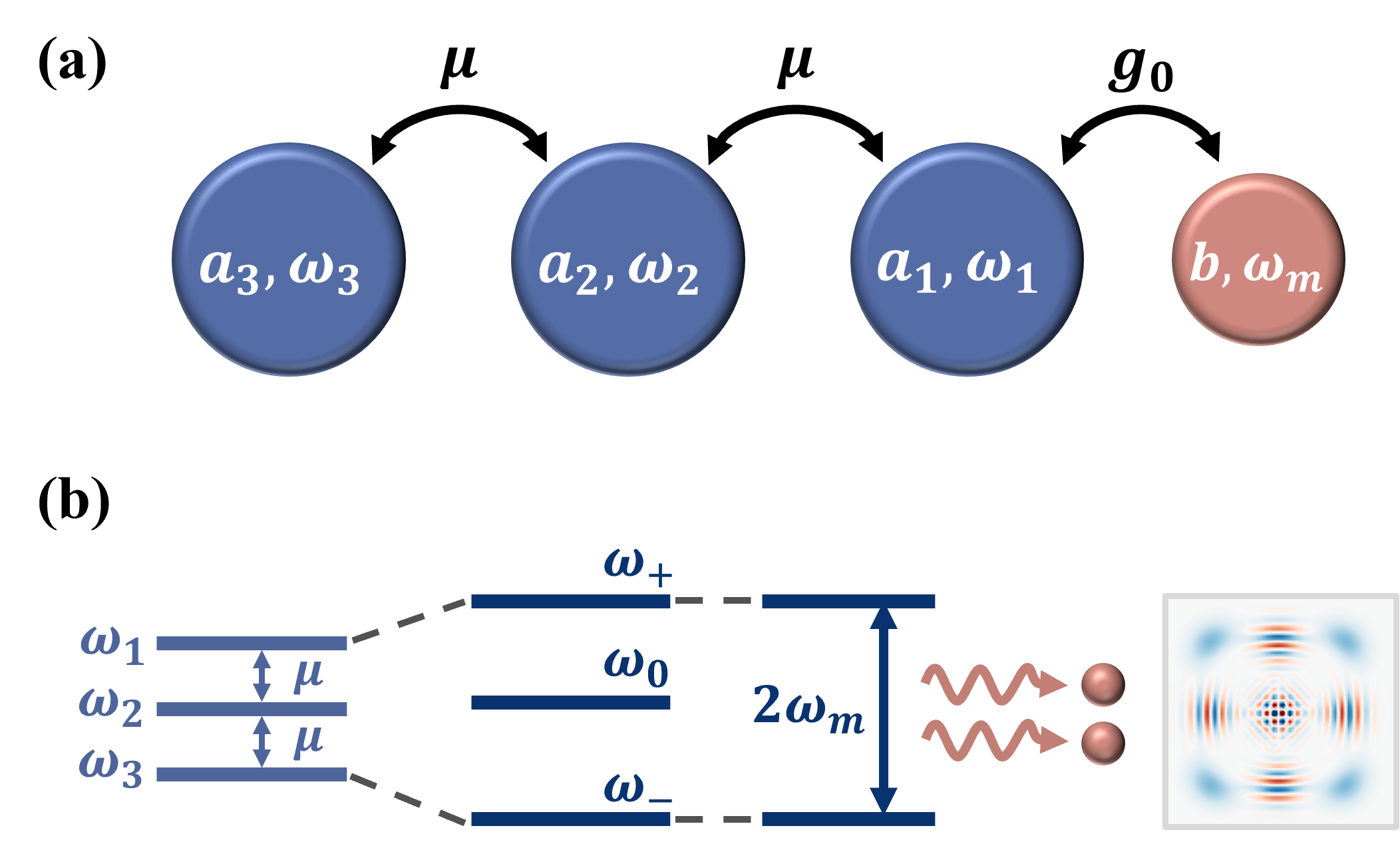}
	\caption{\label{fig:1}(a) Schematic of  multimode optomechanical cavities with three optical modes (in blue) and one mechanical mode (in orange). (b) Schematic of two-phonon resonance driving the generation of multicomponent mechanical cat states.} 
\end{figure}

Here, using only linear optomechanical coupling, we first achieve the two-phonon resonance in a multimode system, where the interaction between photons and two phonons is significantly enhanced.  
As shown in Fig.~\ref{fig:1}, when the frequencies $\omega_{+}$ and $\omega_{-}$ of two optical supermodes satisfy the two-phonon resonance condition, i.e., $\omega_{+}-\omega_{-} \approx 2 \omega_m$, the  two-phonon process mediated by the auxiliary supermode  $\omega_{0}$ is greatly strengthened. 
This process involves the annihilation of a $\omega_{+}$ photon accompanied by the creation of a $\omega_{-}$ photon and two $\omega_m$ phonons, and vice versa. 
Remarkably, the strength of this two-phonon interaction  can reach 1/40 of the linear optomechanical coupling, which far exceeds those in previous studies \cite{2023PRL.HAU,2013PRA.TAN,2010PRA,2018PRA.BRU,2019PRA,2024PRA}. 
Such resonant two-phonon process can drive multiple rotations and interferences of mechanical coherent states in phase space, deterministically generating a MMCS with high purity, as shown in Fig. \ref{fig:1}(b).
In addition, owing to the enhanced interaction strength, the purity of generated states is immune to both mechanical and optical losses. 
Our work shows that introducing multiple optical modes into linear optomechanical system can substantially enhance high-order phonon nonlinearities, opening avenues for quantum state engineering, quantum precision measurement and fault-tolerant quantum computation.
 
\section{Two-Phonon Resonance}

Consider an optomechanical system with three optical modes and a mechanical mode, as shown in Fig.~\ref{fig:1}(a). The optical modes $a_{1},a_{2},a_{3}$ with frequencies $\omega_{1},\omega_{2},\omega_{3}$ are  coupled with strength $\mu$ and the mechanical mode $b$ with frequency $\omega_{m}$ is coupled with $a_{1}$ via radiation pressure with coupling strength $g_{0}$. The system is governed by  the Hamiltonian
\begin{equation}
	\begin{aligned}
		H &= \hbar \omega_1 a_1^{\dagger} a_1 
		+ \hbar \omega_2 a_2^{\dagger} a_2
		+ \hbar \omega_3 a_3^{\dagger} a_3 
		+ \hbar \mu \bigl(a_1^{\dagger} a_2 + a_2^{\dagger} a_1 \bigr)
		+ \hbar \mu \bigl(a_2^{\dagger} a_3 + a_3^{\dagger} a_2 \bigr) \\
		&\quad + \hbar \omega_m b^{\dagger} b
		- \hbar g_0 a_1^{\dagger} a_1 \bigl( b + b^{\dagger} \bigr)\,.
	\end{aligned}
\end{equation}	
To simplify, we assume $\omega_{3}=\omega_{1}$. Since here $\mu \gg g_{0}$,  optical supermodes $a_{+},a_{0},a_{-}$  can be first set up by an unitary transformation
\begin{equation}
	\begin{pmatrix} a_+ \\ a_0 \\ a_- \end{pmatrix}
	= M \begin{pmatrix} a_1 \\ a_2 \\ a_3 \end{pmatrix},
	\label{eq:relation}
\end{equation}
where
\begin{equation}
	M =
	\begin{pmatrix}
		\frac{\sqrt{2}\,\mu}{\sqrt{\Delta(\Delta+\omega_2-\omega_1)}} &
		\frac{1}{\sqrt{2}}\sqrt{1+\frac{\omega_2-\omega_1}{\Delta}} &
		\frac{\sqrt{2}\,\mu}{\sqrt{\Delta(\Delta+\omega_2-\omega_1)}} \\[2mm]
		-\frac{1}{\sqrt{2}} & 0 & \frac{1}{\sqrt{2}} \\[2mm]
		\frac{\sqrt{2}\,\mu}{\sqrt{\Delta(\Delta+\omega_1-\omega_2)}} &
		-\frac{1}{\sqrt{2}}\sqrt{1+\frac{\omega_1-\omega_2}{\Delta}} &
		\frac{\sqrt{2}\,\mu}{\sqrt{\Delta(\Delta+\omega_1-\omega_2)}}
	\end{pmatrix}
	\label{eq:M}
\end{equation}
with the eigenfrequencies $\omega_\pm = \frac{1}{2} \bigl( \omega_1 + \omega_2 \pm \Delta \bigr)$ and $\omega_0 = \omega_1$, in which $\Delta  = \sqrt{8\,\mu^2 + (\omega_1 - \omega_2)^2}$, as shown in Fig.~\ref{fig:1}(b).
Thus, the optomechanical interaction  occurs between the mechanical mode and three optical supermodes. The transformed Hamiltonian in the supermode basis is shown in Supplementary Material \cite{SM}.

\begin{figure}[t]
	\includegraphics[width=0.8\textwidth]{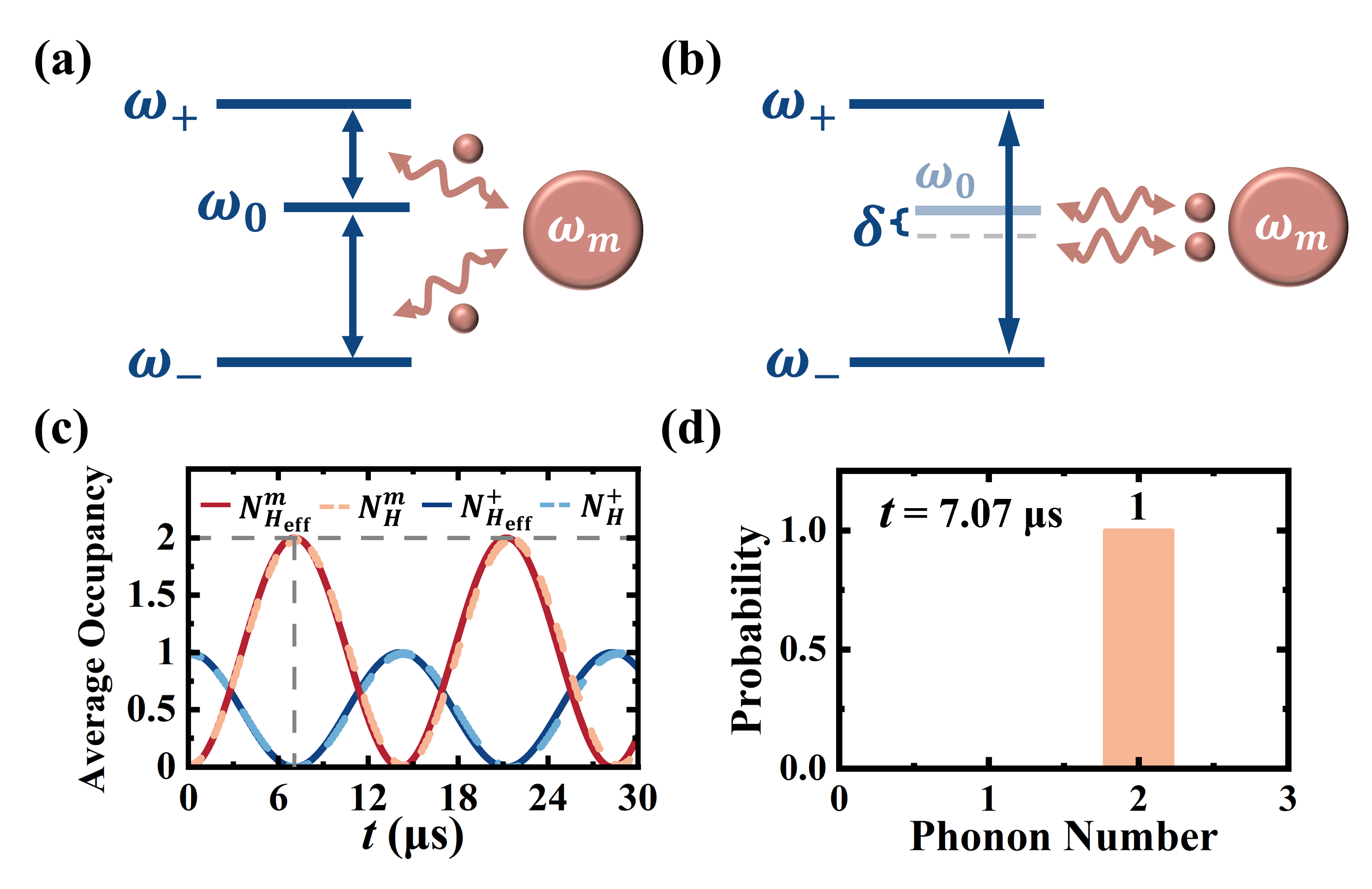}
	\caption{\label{fig:2}(a) Two separate one-phonon process occurring between supermodes $a_{+}$ and $a_{0}$ and  between $a_{0}$ and $a_{-}$. (b) Two-phonon resonance process where an $\omega_{+}$ photon is annihilated with the creation of an $\omega_{-}$ photon and two phonons, and vice versa.
		(c) Two-phonon generation process with original  $H$ and effective  $H_{\text{eff}}$. $N_{H_{\text{eff}}(H)}^{m(+)}$ is the average phonon number (photon number of supermode $a_{+}$) under $H_{\text{eff}}$ ($H$). 
		(d) The phonon number distribution at $t=7.07 \mu s$.  Here, $g_0/2\pi = 1\ \mathrm{MHz}$, $\omega_m/2\pi = 5\ \mathrm{GHz}$, $\delta/2\pi = 5\ \mathrm{MHz}$, $g = g_0/40$, and $\omega_+ - \omega_- - 2\omega_m = -g$.
	} 
\end{figure}

Then, under the condition $g_0, |\delta_1|, |\delta_2| \ll \omega_m$ with the detunings $\delta_1 = \omega_+ - \omega_0 - \omega_m$ and $\delta_2 = \omega_0 - \omega_- - \omega_m$, the Hamiltonian in the rotating wave approximation is given by
\begin{equation}
	H' = \hbar g_1 \bigl( a_+ a_0^\dagger b^\dagger e^{-i \delta_1 t} + a_+^\dagger a_0 b e^{i \delta_1 t} \bigr)
	+ \hbar g_2 \bigl(a_0 a_-^\dagger b^\dagger e^{-i \delta_2 t} + a_0^\dagger a_- b e^{i \delta_2 t}\bigr)
	\label{eq:4}
\end{equation}
with $g_1 \approx g_2 \approx g_0 / (2\sqrt{2})$. It means that  two single-phonon processes occur between $a_+,a_0$ and between $a_0,a_-$, respectively, as shown in Fig.~\ref{fig:2}(a). 
Please note that if  $\omega_1=\omega_2$, then $\delta_1=\delta_2=0$, so the system will behave as two independent single-phonon processes rather than an effective two-phonon process.

Physically, when the single-phonon process are far detuned, i.e., $\lvert\delta_1\rvert \gg g_1,\ \lvert\delta_2\rvert \gg g_2$, a photon transitions from  $a_{+}$ to $a_{0}$ accompanied by the emission of a phonon while this photon cannot be stored in the intermediate supermode $a_0$. Thus, it must immediately jump to  $a_{-}$, resulting in the creation of two phonons in the whole process.
Especially, under two-phonon resonance, i.e., $\omega_+ - \omega_- \approx 2\omega_m$, this process of generating two phonons will be significantly enhanced. 
From Eq. (\ref{eq:4}), when two single-phonon processes are far detuned and simultaneously two-phonon resonance is satisfied, by using time-averaging method \cite{2000FDP,2007CJP}, an effective Hamiltonian is obtained
\begin{equation}
	H_{\mathrm{eff}} = -\hbar g \bigl[ a_+ a_-^\dagger b^{\dagger 2} e^{-i(\omega_+ - \omega_- - 2\omega_m)t} + \text{h.c.} \bigr] - \hbar g \bigl[ a_+^\dagger a_+ + a_+^\dagger a_+ b^\dagger b + a_-^\dagger a_- b^\dagger b \bigr]
	\label{eq:5}
\end{equation}
where $g = g_0^2/8\delta$ with $\delta_2 \approx -\delta_1 \approx \delta = (\omega_1 - \omega_2)/2$. Eq.~(\ref{eq:5}) is one of the main result of this work.
The first two terms describes a two-phonon process between $a_{+}$ and $a_{-}$, as illustrated in Fig.~\ref{fig:2}(b).
While, the terms $a_+^\dagger a_+ b^\dagger b$ and $a_-^\dagger a_- b^\dagger b$  represent Kerr-type nonlinear interactions between photons and phonons.
By decreasing the controllable parameter $\delta$, two-phonon interaction strength $g$ can be significantly enhanced. 
But here $\delta$ cannot be too small, otherwise, the condition $|\delta| \gg g_0$ for Eq.~(\ref{eq:5}) fails. Below, we will show that $\delta$ can be set as low as $5g_0$, yielding an interaction strength of  $g = g_0/40$, which
far exceeds those in previous studies \cite{2013PRA.TAN,2010PRA,2018PRA.BRU,2019PRA,2024PRA,2023PRL.HAU}.
Such superiority comes from the fact that our method utilizes linear optomechanical coupling to obtain  large  $g$,  instead of directly using extremely weak quadratic coupling \cite{2013PRA.TAN,2010PRA,2018PRA.BRU,2019PRA,2024PRA}.
Although the scheme \cite{2023PRL.HAU} is also based on linear optomechanical coupling, its interaction strength is proportional to $g_0^2/\omega_m$, which is typically $g_0/10^4$ to $g_0/10^3$ \cite{2023NC.BUR,2012APL.CHA}.

The two-phonon interaction in $H_{\mathrm{eff}}$ is  different  from those obtained in previous works \cite{2023PRL.HAU,2013PRA.TAN,2010PRA,2018PRA.BRU,2019PRA,2024PRA}, in which the interaction occurs between one optical and one mechanical mode, i.e., a photon is annihilated with the creation of two phonons assisted by a drive field. 
But here, the two-phonon interaction arises  between two optical modes and one mechanical mode, that is, the annihilation of an $\omega_{+}$ photon accompanied by the creation of an $\omega_{-}$ photon and two $\omega_m$ phonons. 
Specially, $H_{\mathrm{eff}}$ exhibits the property of photon number conservation, which ensures the deterministic generation of the multicomponent mechanical cat states as well as  their robustness against optical loss.

To satisfy conditions $g_0 \ll \delta \ll \omega_m$ and $\omega_+ - \omega_- \approx 2\omega_m$,
 we let $g_0/2\pi=1\ \mathrm{MHz}$, $\delta/2\pi = 5\ \mathrm{MHz}$, $\omega_m/2\pi = 5\ \mathrm{GHz}$, and $\omega_+ - \omega_- - 2\omega_m=-g$. With these parameters,  the  strength  $g$ of two-phonon interaction in our scheme can achieve $g_0/40$. 
For an initial state where a photon is in mode $a_+$ while $a_-$ and $b$ are both in vacuum, a pure two-phonon state is generated at $t=7.07\ \upmu\mathrm{s}$ when an $\omega_{+}$ photon transitions to an $\omega_{-}$ photon, as shown in Figs. \ref{fig:2}(c) and (d).
There is a good agreement between the results obtained from $H$ and $H_{\mathrm{eff}}$, which means that set $\delta=5g_0$ is valid for Eq.~(\ref{eq:5}). 
In fact, $g_0/40$ is a conservative estimate. In the Supplemental Material \cite{SM}, we  demonstrate that by further decreasing $\delta$, $g$ can be  increased  to $g_0/8$ for the generation of two-phonon Fock states and to $g_0/18$ for the generation of four-component mechanical cat states. 
Such two-phonon resonance process will be the foundation to generate MMCS.

\section{Generation of multicomponent mechanical cat states }

Through optomechanical nonlinearities \cite{2023PRL.HAU, 2013PRA.TAN, 2018PRA.BRU}, two-component mechanical cat states have been theoretically prepared. Nevertheless, due to the weak two-phonon interaction in these schemes, the generated cat states  will be destroyed by large mechanical decoherence, especially in high-temperature environments.
In the following, based on strong interaction enhanced by two-phonon resonance,  the generated  MMCS is robust against both mechanical and optical decoherence.
Before cat state generation, we eliminate the Kerr terms in Eq.~(\ref{eq:5}) to obtain a pure two-phonon interaction. 
With  $a_+,a_-$ in Fock state $|n0\rangle$, owing to the photon number conservation in $H_{\mathrm{eff}}$, i.e., $a_+^\dagger a_+ + a_-^\dagger a_-$ is always $n$,  the total Kerr term $a_+^\dagger a_+ b^\dagger b+a_-^\dagger a_- b^\dagger b$ reduces to $nb^\dagger b$. 
Then, using the rotating wave transformation (More details are shown in Ref. \cite{SM}) and setting $\omega_+ - \omega_- - 2\omega_m = (1 - 2n)g$ (for not very large $n$, the two-phonon resonance remains), we have
\begin{equation}
	H_{\mathrm{eff}}' = -\hbar g \bigl[ a_+^\dagger a_- b^2 + \text{h.c.} \bigr],
\label{eff'}
\end{equation}
which, only including two-phonon interaction, is the foundation of generating MMCSs.

In principle, with $|\Psi\rangle_{\mathrm{in}} = |n0\rangle |\alpha\rangle$, i.e., $a_+,a_-$ are in Fock state $|n0\rangle$ while $b$ is in a large coherent state $|\alpha\rangle$, $n+1$ component mechanical cat states can be generated through the evolution of this optomechanical system governed by $H_{\mathrm{eff}}'$. Here, we only present the example of  generating 4-component mechanical cat state when $n=3$. 
First, we give the state evolution when optical modes are in an ``eigenstate'' while mechanical mode is in a coherent state.
Due to the photon number conservation, the optical part of the quantum state is confined to the 3-photon subspace, where the four eigenstates and eigenvalues of $-\hbar g \bigl[ a_+^\dagger a_- + \text{h.c.} \bigr]$ are
\begin{equation}
	\begin{gathered}
		|\psi_\pm\rangle = \tfrac{1}{2\sqrt{2}}\bigl( |30\rangle \pm \sqrt{3} |21\rangle + \sqrt{3} |12\rangle \pm |03\rangle \bigr),E_\pm^\psi = \mp 3\hbar g, \\
		|\phi_\pm\rangle = \tfrac{1}{2\sqrt{2}}\bigl( \sqrt{3} |30\rangle \pm |21\rangle - |12\rangle \mp \sqrt{3} |03\rangle \bigr),E_\pm^\phi = \mp \hbar g.
	\end{gathered}
\end{equation}
When the optical modes are in eigenstates, according to the Hamiltonian in Eq. (\ref{eff'}), the quantum states of the system  evolve as 
\begin{equation}
	\begin{aligned}
		|\psi_\pm\rangle |\alpha\rangle &\rightarrow \frac{1}{2\sqrt{2}} \left( e^{\pm i \frac{21}{2} gt} |30\rangle \pm \sqrt{3} e^{\pm i \frac{9}{2} gt} |21\rangle + \sqrt{3} e^{\mp i \frac{3}{2} gt} |12\rangle \pm e^{\mp i \frac{15}{2} gt} |03\rangle \right) \otimes |\alpha e^{\pm i3gt}\rangle, \\
		|\phi_\pm\rangle |\alpha\rangle &\rightarrow \frac{1}{2\sqrt{2}} \left( \sqrt{3} e^{\pm i \frac{7}{2} gt} |30\rangle \pm e^{\pm i \frac{3}{2} gt} |21\rangle - e^{\mp i \frac{1}{2} gt} |12\rangle \mp \sqrt{3} e^{\mp i \frac{5}{2} gt} |03\rangle \right) \otimes |\alpha e^{\pm igt}\rangle.
	\end{aligned}
\label{eigenevol}
\end{equation}
It means that under  two-phonon resonance, for each eigenstate, the mechanical coherent state rotates at an angular velocity proportional to corresponding eigenvalue \cite{SM}. 

Then, for input state $|\Psi\rangle_{\mathrm{in}} = |30\rangle |\alpha\rangle$, where the optical part 
$|30\rangle$ is decomposed into four eigenstates,  according to Eq.~(\ref{eigenevol}), the mechanical coherent state rotates at different angular velocities.
The quantum state at  time $t$ evolves  as
\begin{equation}
	|\Psi(t)\rangle = |30\rangle |\Psi_0\rangle + |21\rangle |\Psi_1\rangle + |12\rangle |\Psi_2\rangle + |03\rangle |\Psi_3\rangle
\end{equation}
with
\begin{equation}
	\begin{aligned}
		|\Psi_0\rangle &= \frac{1}{8} \left( e^{i \frac{21}{2} gt} |\alpha e^{i3gt}\rangle + 3 e^{i \frac{7}{2} gt} |\alpha e^{igt}\rangle + 3 e^{-i \frac{7}{2} gt} |\alpha e^{-igt}\rangle + e^{-i \frac{21}{2} gt} |\alpha e^{-i3gt}\rangle \right), \\
		|\Psi_1\rangle &= \frac{1}{8} \left( \sqrt{3} e^{i \frac{9}{2} gt} |\alpha e^{i3gt}\rangle + \sqrt{3} e^{i \frac{3}{2} gt} |\alpha e^{igt}\rangle - \sqrt{3} e^{-i \frac{3}{2} gt} |\alpha e^{-igt}\rangle - \sqrt{3} e^{-i \frac{9}{2} gt} |\alpha e^{-i3gt}\rangle \right), \\
		|\Psi_2\rangle &= \frac{1}{8} \left( \sqrt{3} e^{-i \frac{3}{2} gt} |\alpha e^{i3gt}\rangle - \sqrt{3} e^{-i \frac{1}{2} gt} |\alpha e^{igt}\rangle - \sqrt{3} e^{i \frac{1}{2} gt} |\alpha e^{-igt}\rangle + \sqrt{3} e^{i \frac{3}{2} gt} |\alpha e^{-i3gt}\rangle \right), \\
		|\Psi_3\rangle &= \frac{1}{8} \left( e^{-i \frac{15}{2} gt} |\alpha e^{i3gt}\rangle - 3 e^{-i \frac{5}{2} gt} |\alpha e^{igt}\rangle + 3 e^{i \frac{5}{2} gt} |\alpha e^{-igt}\rangle - e^{i \frac{15}{2} gt} |\alpha e^{-i3gt}\rangle \right).
	\end{aligned}
\end{equation}
It is seen that $|\Psi_0\rangle$, $|\Psi_1\rangle$, $|\Psi_2\rangle$, $|\Psi_3\rangle$ are all 4-component cat states.
As a result, after performing a photon number measurement, regardless of which state they collapse into, a 4-component mechanical cat state is deterministically generated. Obviously, for $|\Psi\rangle_{\mathrm{in}} = |n0\rangle |\alpha\rangle$, an $n+1$ component cat state can be generated because $n+1$ optical eigenstates induce $n+1$ different rotation velocities \cite{SM}.

\begin{figure}[t]
	\includegraphics[width=0.8\textwidth]{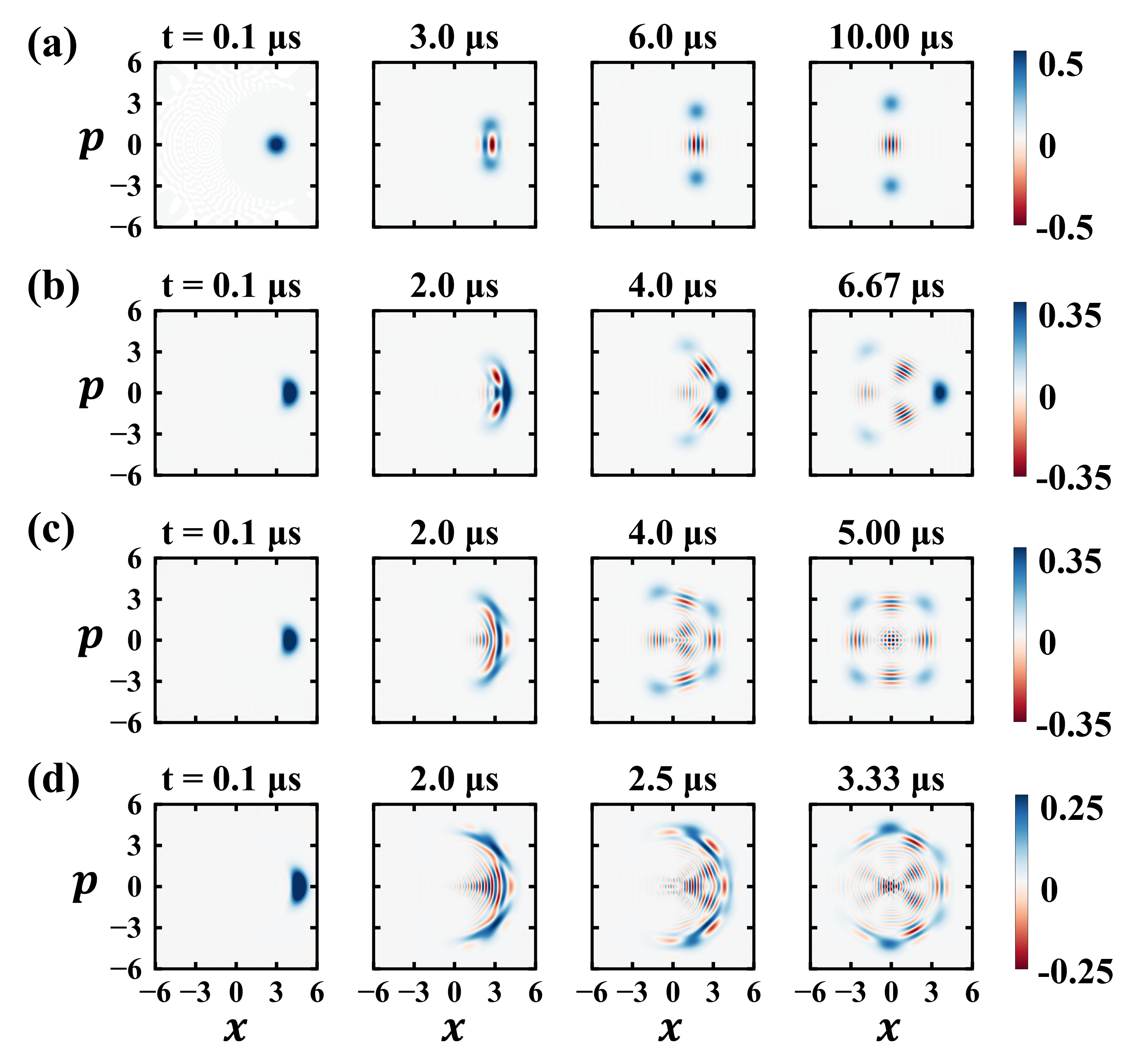}
	\caption{\label{fig:3}The evolution of  Wigner distributions for (a) 2, (b) 3, (c) 4, (d) 6-component cat states after conditional measurement. 
		Here the initial mechanical state is $ |\alpha=3\rangle$ and optical states are  $|10\rangle$ for (a),  $|20\rangle$ for (b), $|30\rangle $ for (c), and $|50\rangle $ for (d) respectively.} 
\end{figure}

Next, based on $H_{\mathrm{eff}}'$ in Eq.~(\ref{eff'}), we numerically demonstrate the generation of MMCSs using QuTiP \cite{2026qutip}. Let $g_0/2\pi = 1\ \mathrm{MHz}$, $\delta/2\pi = 5\ \mathrm{MHz}$, $|\Psi\rangle_{\mathrm{in}} = |n0\rangle |\alpha=3\rangle$, the evolution of mechanical quantum state after conditional measurement is shown in Fig.~\ref{fig:3}, where the  2, 3, 4, and 6-component cat states  appear  when input optical states are $|10\rangle$,  $|20\rangle$,  $|30\rangle$, and $|50\rangle$ respectively. Moreover, due to  phonon addition resulted from the conditional detection, the states generated here are  similar to squeezed cat states, which are more robust to loss and noise \cite{2018PRL.JEA,2023NPJ}.

\begin{figure}[t]
	\includegraphics[width=0.8\textwidth]{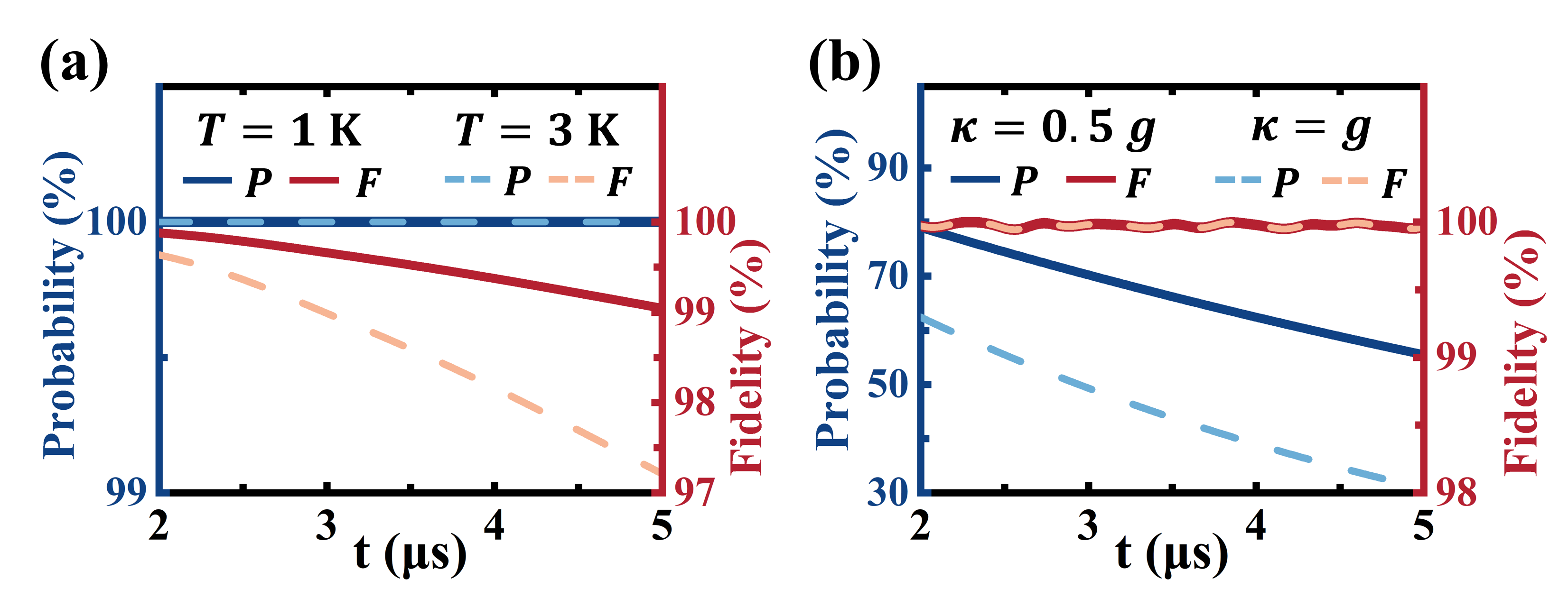}
	\caption{\label{fig:4}The probability and fidelity of the generated states with different (a) thermal noise and (b) optical losses. Here, $\omega_m = 10\ \mathrm{GHz}$, $\gamma = 10\ \mathrm{Hz}$, and the other parameters are the same as those  in Fig.~3.} 
\end{figure}

Using the Lindblad equation, we investigate the influence of mechanical decoherence under realistic conditions ($\omega_m = 10$~GHz, $\gamma = 10$~Hz~\cite{2020NC} and temperature $T \sim$ several K). As illustrated in Fig.~\ref{fig:4}(a), the 4-component cat states are generated with $100\%$ probability and the fidelity with respect to the lossless case remains above $97\%$ over $5$~$\mu$s. Therefore, generation probability and fidelity are almost unaffected by mechanical decoherence, since here $g$ far exceeds $\gamma (n_{\mathrm{th}} + 1)$ with the thermal phonon number $n_{\mathrm{th}} =1\sim5$ at $T$. 
In contrast to previous studies \cite{2023PRL.HAU,2013PRA.TAN,2018PRA.BRU} requiring mK environments, two-phonon resonance here yields several orders higher interaction strength,  so that high-purity cat states can be deterministically prepared even at several K.

Besides, the fidelity of these cat states is highly immune to optical loss. Even when the loss $\kappa$ of mode $a_1$ is comparable to $g$, the fidelity of 4-component cat states is always near 1, indicating that it is still a pure cat state [Fig.~\ref{fig:4}(b)]. Such immunity to loss originates from photon number conservation under the two-phonon resonance \cite{SM}.
In this case, the generation of cat states is no longer deterministic and the  generation probability decreases with the increment of loss.

Finally, we discuss the possibility of experimental realization. The cavity optomechanical system with two optical modes has been fabricated in optomechanical crystals \cite{2023NC.BUR} and whispering gallery microcavities \cite{2010PRL,2018NPHO}. By adding an optical microcavity, our system with three optical modes may be realized. The required conditions $g_0 \ll |\delta| \ll \omega_m$ and $\omega_+ - \omega_- \approx 2\omega_m$ could be achieved by tuning the cavity frequency differences and the optical coupling strength $\mu$. Consequently, it is  expected that realizing two-phonon resonance is experimental accessible. 
The main challenge in generating MMCSs is that the success probability is reduced by optical loss, as the input Fock state is highly sensitive to it. Therefore, the ratio of the optomechanical coupling strength $g_0$ to the optical loss $\kappa$ needs to be substantially improved beyond the current experimental level ($g_0/\kappa \sim 0.01$ \cite{2012APL.CHA}). With further advancements in optomechanical coupling strength and the quality factor of optical microcavities, our scheme may be promising to efficiently generate high-purity MMCSs in the near future.

\section{Conclusion}
Only relying on linear optomechanical coupling, we have achieved the two-phonon resonance in a multimode cavity optomechanical system, where the strength of two-phonon interaction is strongly enhanced to $g_0/40$. Based on this, we have demonstrated the deterministic generation of MMCSs from initial optical Fock state and mechanical coherent state. Specially, the fidelity of the generated states is immune to both optical and mechanical losses. The strong two-phonon nonlinearity achieved here is not only crucial for cat state generation but also has potential for fault-tolerant quantum computation. Building on the strategy established in our work, enhancing three- or four-phonon interactions is possible, thereby offering a framework for engineering optomechanical nonlinearities. With such rich and controllable nonlinearities, our system holds great promise for performing various quantum tasks at the single-quantum level, thus providing a unique platform toward multifunctional quantum processors and quantum network nodes.
\\

\begin{acknowledgments}
This work is supported by the National Natural Science Foundation of China under Grant No. U25D9003  and No. 12474370 and the Quantum Science and Technology-National Science and Technology Major Project No. 2021ZD0301500.
\end{acknowledgments}


\end{document}